\documentclass{pasj00}
\draft

\begin{document}
\SetRunningHead{Author(s) in page-head}{Running Head}
\Received{2009/10/09}
\Accepted{2009/11/27}

\title{Method of the Particle-in-Cell Simulation for the Y-point in the Pulsar Magnetosphere}

\author{Mitsuhiro \textsc{Umizaki} }
\affil{Graduate School of Science and Engineering, Yamagata University\\
1-4-12 Kojirakawa, Yamagata 990-8560}
\email{mumizaki@ksirius.kj.yamagata-u.ac.jp}
\and
\author{Shinpei \textsc{Shibata}}
\affil{Department of Physics, Yamagata University\\
1-4-12 Kojirakawa, Yamagata 990-8560}
\email{shibata@ksirius.kj.yamagata-u.ac.jp}


\maketitle

\begin{abstract}
Recent observations in X-ray and Gamma-ray suggest that the
emission region of the pulsar magnetosphere can be multifold.
In particular, the open-close boundary of the magnetic field,
so-called the Y-point, can be the place where magnetic field
energy converts into the plasma heat and/or flow energy. Here,
we present a new Particle-in-Cell code in axisymmetric geometry,
which can be applied to the Y-point of the pulsar magnetosphere.
The electromagnetic solver is used in the two-dimensional grid 
points with the cylindrical coordinate ($R$, $z$), 
while the particle solver operates in the three-dimensional 
Cartesian coordinate ($x$, $y$, $z$) by use of Buneman-Boris method. The particle motion can be
relativistic. 
The inner boundary conditions are set up to generate rotation
of the magnetosphere in use of the force-free semi-analytic
solution given by Uzdensky (2003, ApJ, 598, 446).

The code has been verified by dispersion relations of all the
wave modes in an electron-positron plasma.

The initial test run is also presented to demonstrate  the Y-shaped
structure at the top of the dead zone on the light cylinder.
We suggest that the structure is variable with quasi-periodicity
with magnetic reconnection and that plasma will be accelerated
and/or heated.
In the time-averaged point of view, 
break up of the ideal-MHD condition takes place in the vicinity
of the Y-point.
\end{abstract}

\section{Introduction}

\begin{figure}[h!]
 \begin{center}
  \FigureFile(75.3mm, 77.0mm){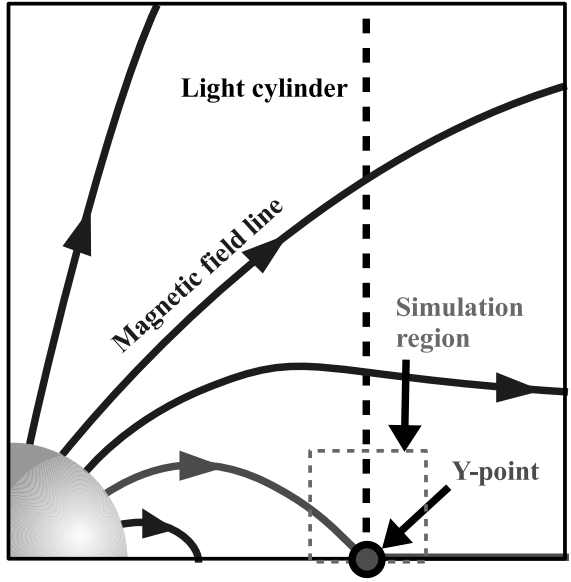}
 \end{center}
 \caption{The meridional plane of the pulsar magnetosphere.}
 \label{magnetosphere}
\end{figure}

Recently many authors have intensively studied the magnetosphere of the
rotation powered pulsars by means of computer simulations in force-free
approximation (\cite{key-9}; \cite{key-10}; \cite{key-11}; \cite{key-5}),
magneto-hydrodynamics (MHD) (\cite{key-1}; \cite{key-8}), two component
MHD (\cite{key-12}) and
particle simulation (\cite{key-6}).  In such studies, boundary layers,
e.g. the equatorial current sheet and the surface of the star, are
sometimes troublesome. Dissipation is suggested in the boundary layers
 (\cite{key-1}), but it can be numerical. Acceleration and heating likely take place in the
magnetic neutral sheet around the equatorial plane. The inner edge of
the neutral sheet, which is called the ``Y-point'', is the
open-close boundary of the magnetic field structure
 (figure \ref{magnetosphere}). Location of the Y-point is not certain
 and can be inside of the light cylinder (\cite{key-2}).

In this paper, we study the Y-point in microscopic point of view to
understand the dissipative process in the boundary layer. We expect that
our simulation provides a hint for the appropriate boundary conditions for
the global simulation. As has been suggested by \citet{key-a1},
another possibility is that the Y-point might be
a gamma-ray source. Pulsed gamma-ray radiation is previously attributed
to the outer gaps and/or the polar caps. One may say that the
radiation near and beyond the light cylinder becomes DC components, but
this may not be the case. Recently, \citet{key-7} proposed that
the magnetic neutral sheet in the wind far from the light cylinder can
be the source of the gamma-ray pulses. If dissipation of the magnetic
 field around the
Y-point takes place, radiation from the vicinity of the Y-point would
also contribute to the pulsed high energy emission.
 
\citet{key-2} calculated the force-free solution around the Y-point. However, he found that the force-free condition is broken around the neutral
sheet. The global MHD simulation \citep{key-1} also suggested breakdown of
force-free condition around the neutral sheet and  the magnetic energy
conversion to the thermal and kinetic energy. The plasma inertia and
magnetic dissipation seem to play an important role around the Y-point.

We intend to study the Y-point via newly developed axisymmetric
 particle-in-cell (PIC) method in the cylindrical coordinates.
 This paper is aimed to establish the numerical code, the
results of code check, and the initialization for the Y-point
 simulation. Some of the initial result shall be given though detailed
 analysis shall be given in a subsequent paper. In our scheme, the
 system is reflection symmetry with respect to the equator,
 which is the lower boundary. The free boundary condition is imposed on the
 outer boundary, which should be continued to the wind
 zone. The left and upper boundaries (inner boundaries) are assumed to
 satisfy the force-free solution given by \citet{key-2}.

\section{Particle-in-Cell Code}

Both particle code and Vlasov code describe the plasma kinetic effect.
However, Vlasov simulation requires huge computer memory, so that we use
the particle code in our simulation. The simplest system to treat the
Y-point would be the axisymmetric one, and therefore we developed
axisymmetric PIC code using cylindrical coordinates.

\subsection{Particle motion solver}

We use Buneman-Boris method in Cartesian coordinates. Then, particles are
distributed in the three-dimensional Cartesian coordinates ($x$,$y$,$z$)
and solved by Buneman-Boris method. Therefore, if a suitable inner
boundary condition of $\boldsymbol{ E}$ and $\boldsymbol{ B}$ are given,
particles rotate around the $z$-axis in the 3D space. Afterwords,
 we transform the particle
coordinates to the cylindrical ones ($R$, $\varphi$, $z$) and solve the
field quantities in the $R$-$z$ plane.

The relativistic equation of motion is
\begin{equation}
 \frac{d \boldsymbol { u}}{d t} = \frac{q}{m} \left[ \boldsymbol { E} + \frac{1}{c\gamma} \boldsymbol { u} \times \boldsymbol { B} \right], \label{eqofm}
\end{equation}
where $\boldsymbol { u}$ is the space part of the four-velocity, $q$ is the charge, $m$ is the rest mass,
$\gamma$ is the Lorentz factor, and $\boldsymbol {E}$ and $\boldsymbol
{B}$ are the electric and the magnetic fields, respectively. In
Cartesian coordinate, $\boldsymbol {E}$ and $\boldsymbol{B}$ are given by
\begin{eqnarray}
 \boldsymbol { E} &=& ( E_R \cos \varphi - E_{\varphi} \sin \varphi, E_R \sin
  \varphi + E_{\varphi} \cos \varphi, E_z ),\\
 \boldsymbol { B} &=& ( B_R \cos \varphi - B_{\varphi} \sin \varphi, B_R \sin \varphi + B_{\varphi} \cos \varphi, B_z ).
\end{eqnarray}
Equation (\ref{eqofm}) is rewritten in a centered-difference form as follows
\begin{equation}
 \frac{\boldsymbol { u}^{n+1/2} - \boldsymbol { u}^{n-1/2}}{\Delta t} = \frac{q}{m} 
 \left(\boldsymbol { E}^n + \frac{\boldsymbol { u}^{n+1/2} + \boldsymbol { u}^{n-1/2}}{2c\gamma}
 \times \boldsymbol { B}^n \right), \label{d_eqofm}
\end{equation}
where $n$ and $\Delta t$ are the number and increment of the time steps.

Buneman-Boris method divides equation (\ref{d_eqofm}) into three steps.

1st step : an acceleration by $\boldsymbol { E}$ for $\Delta t / 2$
\begin{equation}
 \boldsymbol { u}^- = \boldsymbol { u}^{n-1/2} + \frac{q}{m} \boldsymbol { E}^n \frac{\Delta t}{2}.
\end{equation}

2nd step : a rotation by $\boldsymbol { B}$ for $\Delta t$
\begin{eqnarray}
 \boldsymbol { u}^0 &=& \boldsymbol { u}^- + \boldsymbol { u}^- \times \boldsymbol { T}, \\
 \boldsymbol { u}^+ &=& \boldsymbol { u}^- + \frac{2}{1+T^2}\boldsymbol { u}^0 \times \boldsymbol { T},
\end{eqnarray}
where $\boldsymbol { T} = q\boldsymbol { B}^n \Delta t / (2mc\gamma^-)$ and $\gamma^- = \sqrt{1+(u^-/c)^2}$.

3rd step : an acceleration by $\boldsymbol { E}$ for $\Delta t / 2$
\begin{equation}
 \boldsymbol { u}^{n+1/2} = \boldsymbol { u}^+ + \frac{q}{m} \boldsymbol { E}^n \frac{\Delta t}{2}.
\end{equation}
Finally, $\boldsymbol { v}^{n+1/2}$ is calculated as follows,
\begin{equation}
 \boldsymbol { v}^{n+1/2}=\boldsymbol { u}^{n+1/2} \Big/ \sqrt{1+(u^{n+1/2}/c)^2}.
\end{equation}
The position of the particles is updated by
\begin{equation}
 \boldsymbol { x}^{n+1} = \boldsymbol { x}^{n+1/2} + \boldsymbol { v}^{n+1/2} \frac{\Delta t}{2}.
\end{equation}

\subsection{Electromagnetic field solver}

Maxwell equations  are solved in the
cylindrical coordinates with axisymmetry by the Leap-Frog method.
The magnetic field is updated by the Faraday's law,
\begin{equation}
 \frac{\partial \boldsymbol { B}}{\partial t} = - c \nabla \times \boldsymbol { E} \label{fara}
\end{equation}
 and the
electric field is updated by the Amp$\grave{\mathrm{e}}$re's law,
\begin{equation}
  \frac{\partial \boldsymbol { E}}{\partial t} = - 4 \pi \boldsymbol {
  J} + c \nabla \times \boldsymbol { B}, \label{ampere}
\end{equation}
where $\boldsymbol {J}$ is the current density.

\subsubsection{Field definition}

\begin{figure}[h!]
 \begin{center}
  \FigureFile(71.1mm, 64.8mm){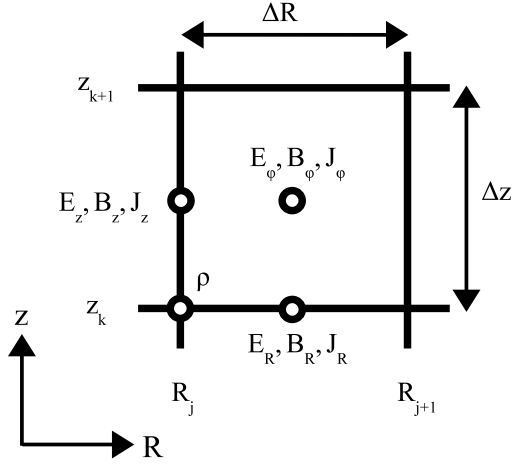}
 \end{center}
 \caption{Location of the grid quantities.}
 \label{teigi}
\end{figure}

In the $R$-$z$ plane, fields are defined as shown in figure \ref{teigi},
with indices $j$ and $k$ for $R$ and $z$ coordinates, respectively. For
example, the $\varphi$-components of $\boldsymbol { E}$, $\boldsymbol { B}$
and $\boldsymbol { J}$ are given at ($R_{j+1/2}, z_{k+1/2}$), but the
$R$-components are given at ($R_{j+1/2}, z_k$).

\subsubsection{Update of magnetic field}

We update the magnetic field by $\Delta t/2$ using the Faraday's law,
\begin{eqnarray}
 \frac{\partial B_R}{\partial t} &=& c \frac{\partial E_{\varphi}}{\partial
 z}, \label{f1}\\
 \frac{\partial B_{\varphi}}{\partial t} &=& - c \left( \frac{\partial E_R}{\partial z} - \frac{\partial E_z}{\partial R}  \right), \label{f2}\\
 \frac{\partial B_z}{\partial t} &=& - \frac{c}{R} \frac{\partial}{\partial R} (R E_{\varphi}),\label{f3}
\end{eqnarray}
which are written in centered-difference form,
\begin{eqnarray}
 &&B^n_{R, j+1/2, k} = B^{n-1/2}_{R, j+1/2, k} \nonumber\\
 &&\hspace{0.5cm}
 +c \frac{E^n_{\varphi, j+1/2, k+1/2} - E^n_{\varphi, j+1/2, k-1/2}}{\Delta z}
\frac{\Delta t}{2}, \label{ff1}\\
 &&B^n_{\varphi, j+1/2, k+1/2} = B^{n-1/2}_{\varphi, j+1/2, k+1/2}
 \nonumber\\
 &&\hspace{0.5cm}
 - c \left( \frac{ E^n_{R, j+1/2, k+1} - E^n_{R, j+1/2, k} }{\Delta z} -
 \frac{ E^n_{z, j+1, k+1/2} - E^n_{z, j, k+1/2} }{\Delta R} \right)
\frac{\Delta t}{2}, \label{ff2}\\
 &&B^n_{z, j, k+1/2} = B^{n-1/2}_{z, j, k+1/2} \nonumber\\
 &&\hspace{0.5cm}
 - \frac{c}{R_j} \frac{R_{j+1/2} E^n_{\varphi, j+1/2, k+1/2} - R_{j-1/2}
 E^n_{\varphi, j-1/2, k+1/2}}{\Delta R} \frac{\Delta t}{2}.\label{ff3}
\end{eqnarray}

\subsubsection{Update of electric field}

We update the electric field by $\Delta t$ using the Amp$\grave{\mathrm{e}}$re's law,
\begin{eqnarray}
 \frac{\partial E_R}{\partial t} &=& -4 \pi J_R - c 
\frac{\partial B_{\varphi}}{\partial z}, \label{A1}\\
 \frac{\partial E_{\varphi}}{\partial t} &=& -4 \pi J_{\varphi} + c
\left( \frac{\partial B_R}{\partial z}-
\frac{\partial B_z}{\partial R} \right), \label{A2}\\
 \frac{\partial E_z}{\partial t} &=& -4 \pi J_z + \frac{c}{R}
\frac{\partial}{\partial R} (R B_{\varphi}), \label{A3}
\end{eqnarray}
which are rewritten in centered-difference form,
\begin{eqnarray}
  &&E^{n+1}_{R, j+1/2, k} = E^n_{R, j+1/2, k} \nonumber\\
  &&\hspace{0.5cm}
 -\left( 4\pi J^{n+1/2}_{R, j+1/2, k} + c
 \frac{B^{n+1/2}_{\varphi, j+1/2, k+1/2} -
 B^{n+1/2}_{\varphi, j+1/2, k-1/2}}{\Delta z} \right) \Delta t,\label{AA1} \\
  &&E^{n+1}_{\varphi, j+1/2, k+1/2} = E^n_{\varphi, j+1/2, k+1/2}
  \nonumber\\
  &&\hspace{0.5cm}+\left[ - 4 \pi J^{n+1/2}_{\varphi, j+1/2, k+1/2} +
 c \left( \frac{ B^{n+1/2}_{R, j+1/2, k+1} - B^{n+1/2}_{R, j+1/2,
    k}}{\Delta z} - \frac{B^{n+1/2}_{z, j+1, k+1/2} -
 B^{n+1/2}_{z, j, k+1/2}}{\Delta R} \right) \right] \Delta t,\nonumber \\
 &&\label{AA2}\\
 &&E^{n+1}_{z, j, k+1/2} = E^n_{z, j, k+1/2} \nonumber\\
 &&\hspace{0.5cm}
 + \left( -4 \pi J^{n+1/2}_{z, j, k+1/2} + \frac{c}{R_j}
 \frac{R_{j+1/2} B^{n+1/2}_{\varphi, j+1/2, k+1/2} -
 R_{j-1/2} B^{n+1/2}_{\varphi, j-1/2, k+1/2}}{\Delta R} \right) \Delta t.\label{AA3}
\end{eqnarray}

\subsection{Calculation of charge density}

\begin{figure}[h!]
 \begin{center}
  \FigureFile(65.8mm, 56.7mm){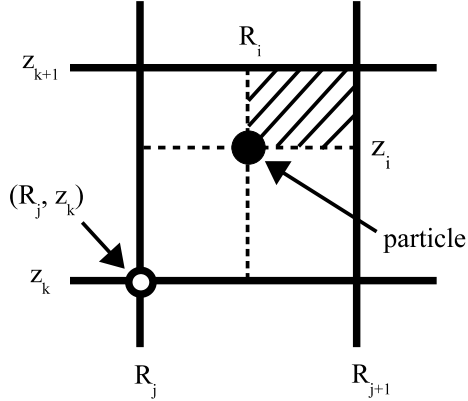}
 \end{center}
 \caption{Schematic diagram for interpolation between a particle and grids.}
 \label{definition}
\end{figure}

To obtain the charge density at ($R_j, z_k$), charge of a particle is
distributed to the neighboring four grids with weightings in proportion to
the volumes opposite to the particle, i.e. the shaded region in figure
\ref{definition} for the left-bottom grid point.
We calculate thus the charge density with shape factors,
\begin{eqnarray}
 S_j \left( R_i \right) = \left\{
      \begin{array}{ll}
       \left( R_{j+1} - R_i \right) / \Delta R &
	\mathrm{for}\hspace{0.2cm} R_j \le R_i < R_{j+1} \\
       \left( R_i - R_{j-1} \right) / \Delta R &
	\mathrm{for}\hspace{0.2cm} R_{j-1} \le R_i < R_j \\
       0 & \mathrm{otherwise} \\
      \end{array} \right.\label{sj}
\end{eqnarray}

\begin{eqnarray}
 S_k \left( z_i \right) = \left\{
      \begin{array}{ll}
       \left( z_{k+1} - z_i \right) / \Delta z &
	\mathrm{for}\hspace{0.2cm} z_k \le z_i < z_{k+1} \\
       \left( z_i - z_{k-1} \right) / \Delta z &
	\mathrm{for}\hspace{0.2cm} z_{k-1} \le z_i < z_k \\
       0 & \mathrm{otherwise} \\
      \end{array} \right.\label{sk}
\end{eqnarray}
in our code.
The charge distributed to ($R_j, z_k$) is
\begin{equation}
 Q_{j, k} = \sum_i q_i  S_j \left( R_i \right) S_k \left( z_i \right),
\end{equation}
which yields the charge density,
\begin{equation}
 \rho_{j,k} = Q_{j,k} / V_{j,k},
\end{equation}
where $V_{j,k} = 2 \pi R_j \Delta R \Delta z$.

On the other hand, the electric field at the position of the $i$-th
 particle ($R_i, z_i$) is given by 
\begin{eqnarray}
 \boldsymbol { E}(R_i, z_i) &=&
  S_j S_k \left(
	   \begin{array}{ccc}
	    (R_j/R_i) E_{R,j,k}\\ 
	    E_{\varphi,j,k}\\
	    E_{z,j,k}\\
	   \end{array}
	  \right) +
  S_j S_{k+1} \left(
	   \begin{array}{ccc}
	    (R_j/R_i) E_{R,j,k+1}\\ 
	    E_{\varphi,j,k+1}\\
	    E_{z,j,k+1}\\
	   \end{array}
	  \right) +\nonumber\\
  &&S_{j+1} S_k \left(
	   \begin{array}{ccc}
	    (R_{j+1}/R_i) E_{R,j+1,k}\\ 
	    E_{\varphi,j+1,k}\\
	    E_{z,j+1,k}\\
	   \end{array}
	  \right) +
  S_{j+1} S_{k+1} \left(
	   \begin{array}{ccc}
	    (R_{j+1}/R_i) E_{R,j+1,k+1}\\ 
	    E_{\varphi,j+1,k+1}\\
	    E_{z,j+1,k+1}\\
	   \end{array}
	  \right)
\end{eqnarray}
The magnetic field is also given in the same form.

\subsection{Calculation of current density}

Current density is also calculated with the shape factors:
\begin{equation}
 \boldsymbol{ J}_{j, k} = \sum_i q_i  S_j
 \left( R_i \right) S_k \left( z_i \right) \boldsymbol{ v}_i.
\end{equation}

We interpolate the current densities at the full integer grids to those
at the appropriate grid points shown in figure \ref{teigi} by
\begin{eqnarray}
 J_{R, j+1/2, k} &=& \frac{R_{j+1} J_{R, j+1,k} + R_j
 J_{R, j,     k}}{2 R_{j+1/2}},\\
 J_{\varphi, j+1/2, k+1/2} &=& \frac{
 J_{\varphi, j+1, k+1} +
 J_{\varphi, j,   k+1} +
 J_{\varphi, j+1, k} +
 J_{\varphi, j,   k}          }{4},\\
 J_{z, j, k+1/2} &=& \frac{J_{z, j, k+1} +
 J_{z, j, k}}{2}.
\end{eqnarray}

Since the above current density does not guarantee the charge conservation law, the electric fields contain an error. Therefore, we use the following
procedure to correct the electric field.
Gauss' law yields
\begin{equation}
 \nabla \cdot \left( \boldsymbol{ E}^{\prime}+ \boldsymbol{ E}_{\mathrm{c}} \right)
  = 4 \pi \rho, \label{gauss}
\end{equation}
where $\boldsymbol{ E}^{\prime}$ is the obtained field by (\ref{AA1}),
(\ref{AA2}), (\ref{AA3}) and $\boldsymbol{ E}_{\mathrm{c}}$ is the correction.
It follows from (\ref{gauss}) that $\boldsymbol{ E}_{\mathrm{c}}$ is obtained by 
\begin{equation}
 \nabla \cdot \boldsymbol{ E}_{\mathrm{c}} = 4 \pi \rho_{\mathrm{c}},
\end{equation}
where $\rho_{\mathrm{c}} = \rho - \nabla \cdot \boldsymbol{ E}^{\prime} / 4 \pi$.
Then, we solve
\begin{equation}
 - \nabla^2 \phi_{\mathrm{c}} = 4 \pi \rho_{\mathrm{c}}
\end{equation}
and we get $\boldsymbol{ E}_{\mathrm{c}} = - \nabla \phi_{\mathrm{c}}.$

\subsection{A calculation of the PIC method for one time step}

\begin{figure*}[h!]
 \begin{center}
  \FigureFile(134.1mm, 51.5mm){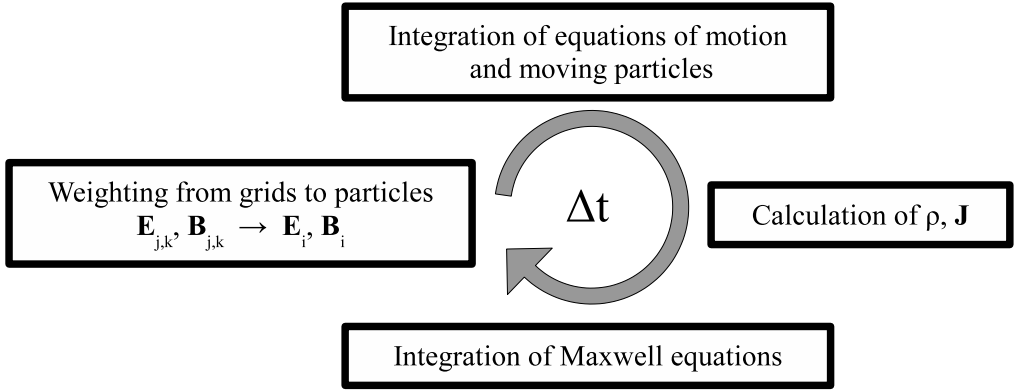}
 \end{center}
 \caption{A calculation cycle of the PIC method.}
 \label{cycle}
\end{figure*}

A calculation cycle of the PIC method is shown in figure \ref{cycle}.
 We interpolate the electromagnetic fields at the particle
positions, $\boldsymbol {E}_i$, $\boldsymbol {B}_i$, from the values of
 the given grids points, $\boldsymbol {E}_{j,k}$, $\boldsymbol
 {B}_{j,k}$.
 We integrate
the equations of motion and move the particle positions. We next
calculate the charge densities and current densities using the particle
 data. Electromagnetic fields are updated by integrating Maxwell
 equations.

\section{Result}

\subsection{Code check}

In order to check the code, we obtain the dispersion
relations of waves in electron-positron plasmas for Light mode, X-mode,
 Fast mode, O-mode, Electromagnetic mode, and Alfv$\acute{\mathrm{e}}$n
 mode. The boundary conditions are perfect conductor for the outer
 radius and periodic for $z$-direction. The plasma is uniform and has
 no bulk velocity. Particle distribution for the thermal velocities is
non-relativistic Maxwellian. The simulation settings are given in
 table \ref{tab:first}.

\begin{table*}[h!]
 \caption{The simulation settings for the code check. $\boldsymbol { B}_0$ is
 background magnetic field, $v_{\mathrm{th}}$ is thermal velocity, $\omega_{\mathrm{p}}$ is plasma frequency, and
 $\Omega$ is gyro-frequency.}\label{tab:first}
 \begin{center}
  \begin{tabular}{|l|l|l|l|}
   \hline
   & Case A
   & Case B
   & Case C \\
   \hline
   $\boldsymbol { B}_0$
   & (0, 0, 0) & (0, $B_{\varphi}$, 0) & (0, 0, $B_z$) \\
   \hline
   $v_{\mathrm{th}}$
   & 0.1$c$ & 0.1$c$ & 0.1$c$ \\
   \hline
   $\Delta R, \Delta z$
   & $v_{\mathrm{th}}\omega_{\mathrm{p}}^{-1}$
   & $v_{\mathrm{th}}\omega_{\mathrm{p}}^{-1}$
   & $v_{\mathrm{th}}\omega_{\mathrm{p}}^{-1}$ \\
   \hline
   $\Delta t$/$\Delta R$
   & 0.2 & 0.2 & 0.2 \\
   \hline
   Grid number ($R \times z$)
   & 128 $\times$ 512
   & 128 $\times$ 512 (X,Fast)
   & 128 $\times$ 512 \\
   && 256 $\times$ 512 (O) &\\
   \hline
   Total time step
   & 4096 & 4096 & 4096 \\
   \hline
   Particle number per grid
   & 200 & 200 & 200 \\
   \hline
   $\Omega / \omega_{\mathrm{p}}$
   & 0 & 1 & 1 \\
   \hline
  \end{tabular}
 \end{center}
\end{table*}

\subsubsection{Light mode}

The simulation for Case A in Table \ref{tab:first} demonstrates the
 light-mode.

The linear dispersion relation of the light mode is
\begin{equation}
 \omega^2 = \omega_{\mathrm{p}}^2 + c^2 k^2, \label{light_mode}
\end{equation}
where
\begin{equation}
 \omega_{\mathrm{p}} = \sqrt{\frac{8\pi n e^2}{m \gamma_{\mathrm{th}}}},
\end{equation}
where
\begin{equation}
 \gamma_{\mathrm{th}} = 1 + \frac{\Gamma}{\Gamma -1} \frac{k_B T}{mc^2},
\end{equation}
$\Gamma$ is a ratio of specific heat, $k_B$ is the Boltzmann
constant, and $T$ is the temperature. The frequency of the light mode
 is $\omega_{\mathrm{p}}$ at $k=0$ and asymptotically
it approaches to $\omega=ck$ in high wave numbers.

\begin{figure}[h!]
 \begin{center}
  \FigureFile(70.0mm, 70.0mm){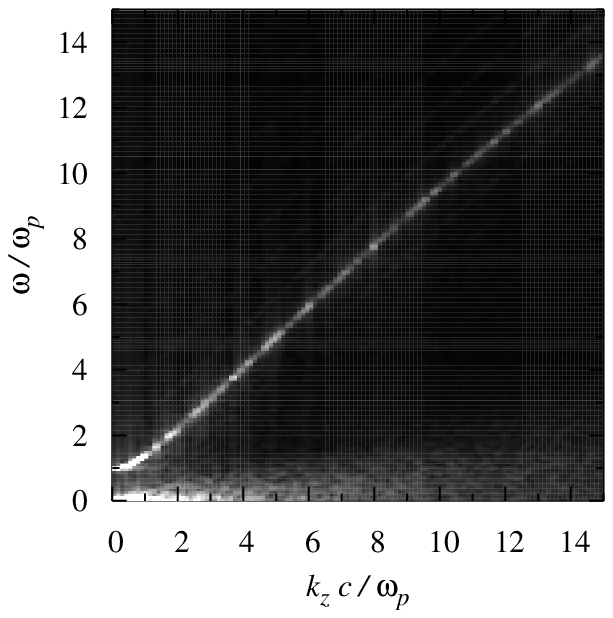}
  \FigureFile(70.0mm, 70.0mm){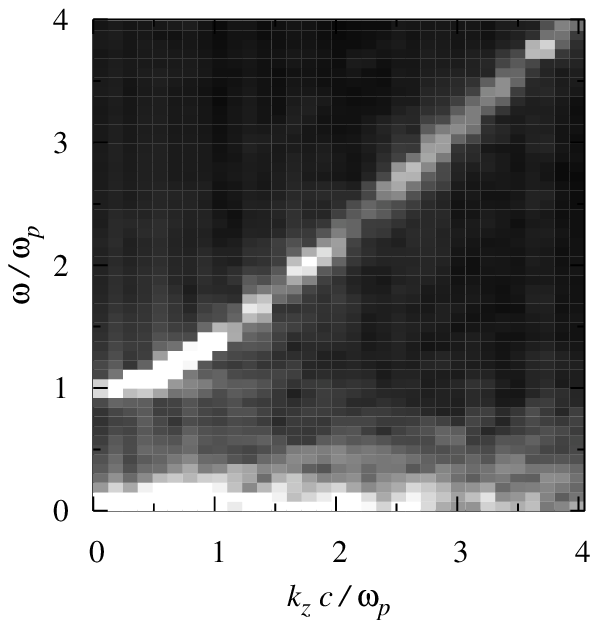}
 \end{center}
 \caption{The dispersion relation diagrams of the light
 mode. Gray map indicates Fourier transformation of $B_{\varphi}$.
 The left figure is the full simulation scale and the right
 is the lower frequency part.}
 \label{lmode_f}
\end{figure}

We compute the Fourier transform with respect to $z$ and then integrate
 that in $R$-direction. We do not impose any kind of initial
 perturbation, i.e., there is only thermal noise initially.
 Figure \ref{lmode_f} shows the result for
 $B_{\varphi}$, showing that dispersion relation (\ref{light_mode}) is well
 reproduced.

\subsubsection{X-mode, Fast mode, and O-mode}

We next analyze the perpendicular propagating waves.
The simulation settings are given in Case B in Table \ref{tab:first}.

The linear dispersion relations of X-mode (positive-signed) and
Fast mode (negative-signed) are
\begin{equation}
 \omega^2 = \frac{1}{2} \left[ (c^2 + c_s^2)k^2 + \Omega^2 +
                         \omega_{\mathrm{p}}^2 \pm \sqrt{ (c^2 - c_s^2)^2 k^4 + 2
                         (c^2 - c_s^2)(\omega_{\mathrm{p}}^2 - \Omega^2)k^2 +
                         (\Omega^2 + \omega_{\mathrm{p}}^2)^2 } \right], \label{xf_mode}
\end{equation}
where
\begin{equation}
\Omega = \frac{eB}{mc\gamma_{\mathrm{th}}},
\end{equation}
and $c_s$ is the speed of the sonic wave,
\begin{equation}
c_s = \sqrt{(\Gamma - 1) \frac{\gamma_{\mathrm{th}} - 1}{\gamma_{\mathrm{th}}}}c.
\end{equation}

The frequency of the X-mode becomes the upper hybrid frequency $\omega_{\mathrm{UH}}=\sqrt{\omega_{\mathrm{p}}^2 +
\Omega^2}$ at $k=0$, and asymptotically approaches $\omega = ck$ in high wave numbers. The frequency of the Fast mode is
 $\omega= 0$ at $k=0$, and asymptotically approaches to $\omega = c_s k$
 in high wave numbers.

\begin{figure}[h!]
 \begin{center}
  \FigureFile(70.0mm, 70.0mm){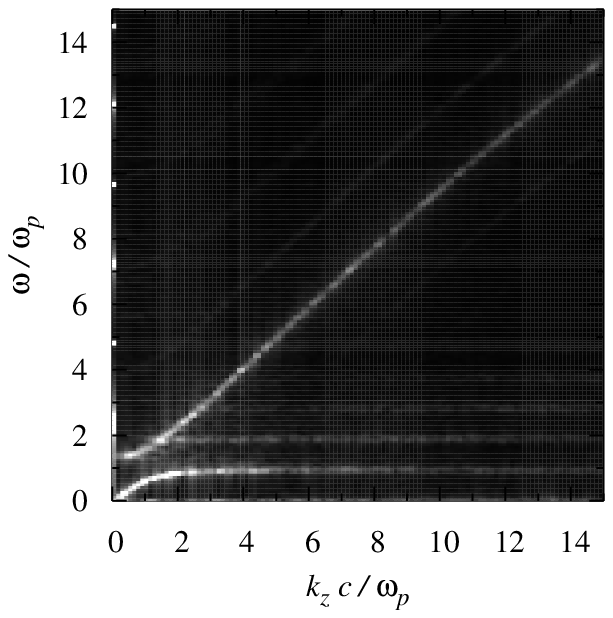}
  \FigureFile(70.0mm, 70.0mm){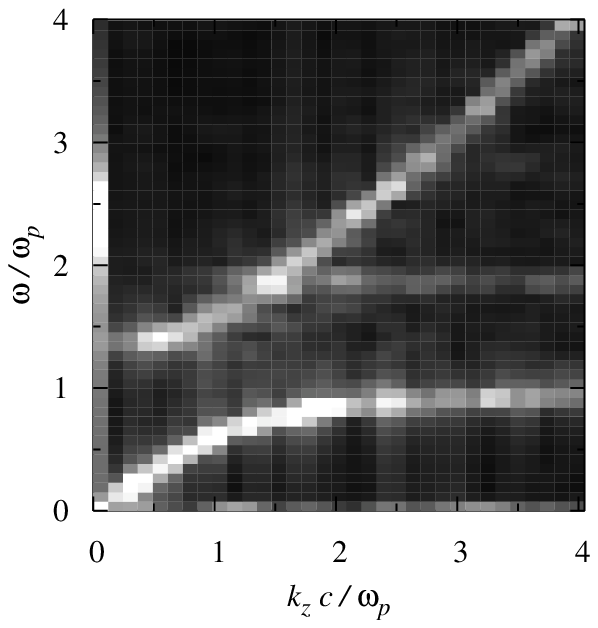}
 \end{center}
 \caption{Fourier transformation of $B_{\varphi}$. The dispersion relation diagrams of the X-mode
 (upper frequency wave) and the Fast mode (lower frequency wave).
 The left figure is the full simulation scale and the right is the
 lower frequency part. For this case, $\omega_{\mathrm{UH}} = \sqrt{2}$ and $c_s
 \approx 6.6 \times 10^{-2}$.}
 \label{fmode_f}
\end{figure}

Figure \ref{fmode_f} shows Fourier transform of
 $B_{\varphi}$, showing that these waves are well fitted by the dispersion
 relations (\ref{xf_mode}).

The linear dispersion relation of O-mode is
\begin{equation}
 \omega^2 = \omega_{\mathrm{p}}^2 + c^2 k^2. \label{o_mode}
\end{equation}

The frequency of the O-mode is $\omega_{\mathrm{p}}$ at $k=0$, and asymptotically
approaches to $\omega = c k$ for high wave numbers.

\begin{figure}[h!]
 \begin{center}
  \FigureFile(70.0mm, 70.0mm){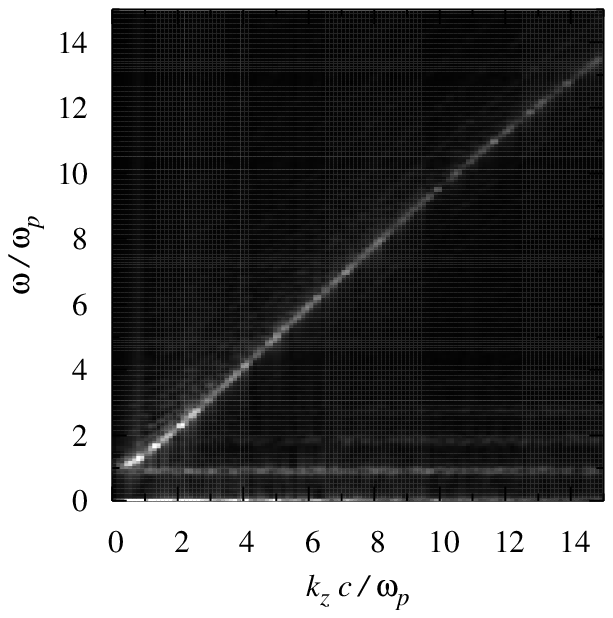}
  \FigureFile(70.0mm, 70.0mm){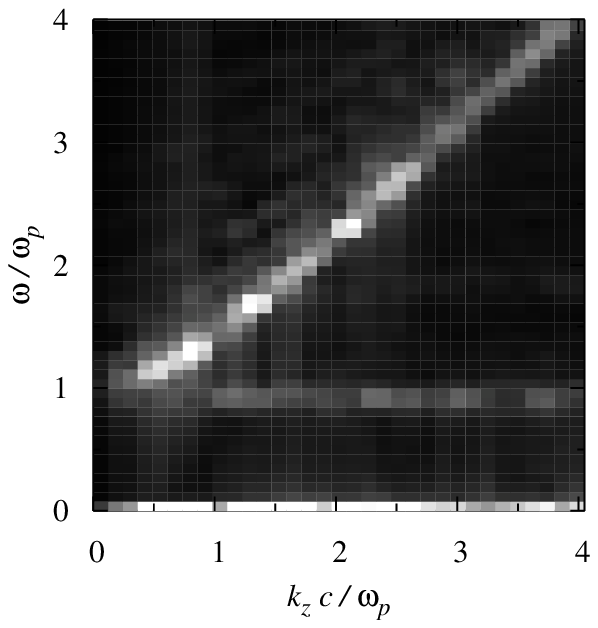}
 \end{center}
 \caption{Fourier transformation of $B_R$ for Case B. The dispersion
 relation of the O-mode is well reproduced. The left figure is the full
 simulation scale and the right is the lower frequency part.}
 \label{omode_f}
\end{figure}

Figure \ref{omode_f} shows Fourier transform of $B_R$, showing that this wave is well fitted by the dispersion relation (\ref{o_mode}).

\subsubsection{Electromagnetic mode and Alfv$\acute{e}$n mode}

Finally, we analyze the parallel propagating waves, Case C, for which
the simulation settings are given in Table \ref{tab:first}.

The linear dispersion relations of the electromagnetic mode (positive-signed) and
Alfv$\acute{\mathrm{e}}$n mode (negative-signed) are
\begin{equation}
 \omega^2 = \frac{1}{2} \left[ c^2 k^2 + \omega_{\mathrm{UH}}^2 \pm
                         \sqrt{ (c^2 k^2 + \omega_{\mathrm{UH}}^2)^2 -
                 4 c^2 k^2 \Omega^2 } \right]. \label{ea_mode}
\end{equation}

The frequency of the electromagnetic mode is $\omega_{\mathrm{UH}}$ at $k=0$, and asymptotically approaches to $\omega=ck$ in high wave numbers. The frequency of the Alfv$\acute{\mathrm{e}}$n mode is $\omega= 0$ at $k=0$, and asymptotically approaches to the relativistic gyro-frequency in high wave numbers.

\begin{figure}[h!]
 \begin{center}
  \FigureFile(70.0mm, 70.0mm){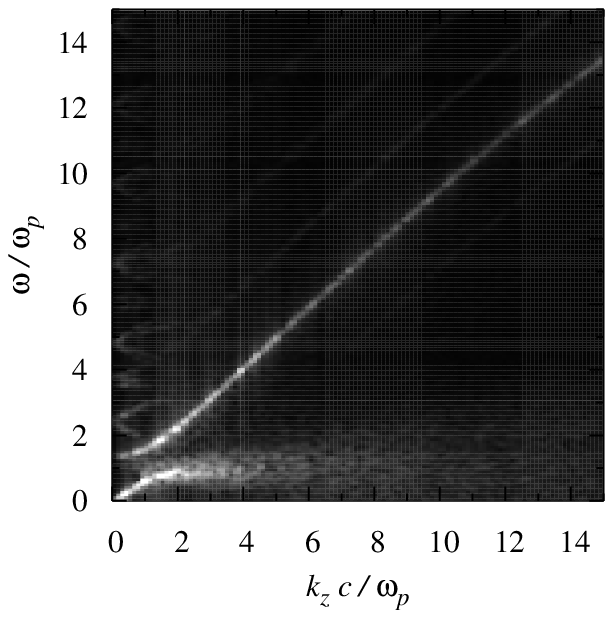}
  \FigureFile(70.0mm, 70.0mm){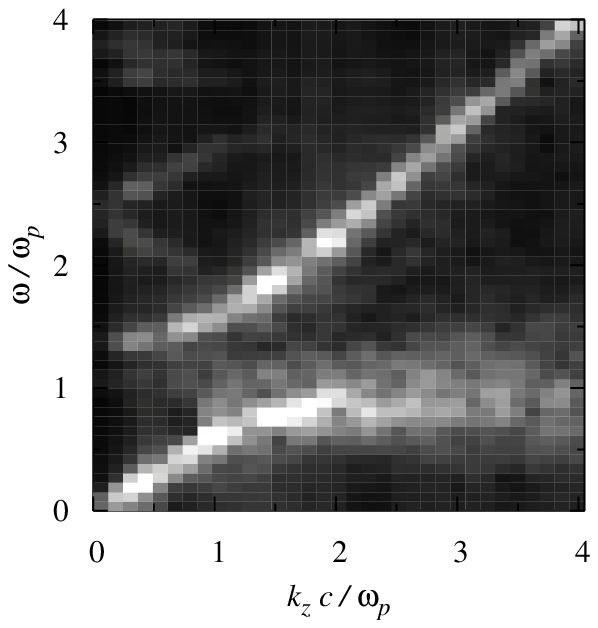}
 \end{center}
 \caption{Fourier transformation of $B_{\varphi}$ for Case C. The
 dispersion relation of the Electromagnetic mode (upper frequency wave)
 and the Alfv$\acute{\mathrm{e}}$n mode (lower frequency wave) can be seen.
 The left figure is the full simulation scale and the right is the lower
 frequency part.}
 \label{amode_f}
\end{figure}

Figure \ref{amode_f} shows Fourier transform of
 $B_{\varphi}$, showing that these waves are well fitted by the dispersion
 relations (\ref{ea_mode}).

\subsubsection{Summary of the code check}

We perform the simulations for waves in electron-positron plasmas in
 order to check the newly developed axisymmetric PIC code.
 We find that dispersion relations of Light mode, X-mode, Fast mode,
 O-mode, Electromagnetic mode, and Alfv$\acute{\mathrm{e}}$n mode are well
 reproduced. Thus we conclude that the code is successfully constructed.

\subsection{Application to the Y-point of the pulsar magnetosphere}

\subsubsection{Outline of the Y-point simulation}

\begin{figure*}[h!]
 \begin{center}
  \FigureFile(139.3mm, 64.8mm){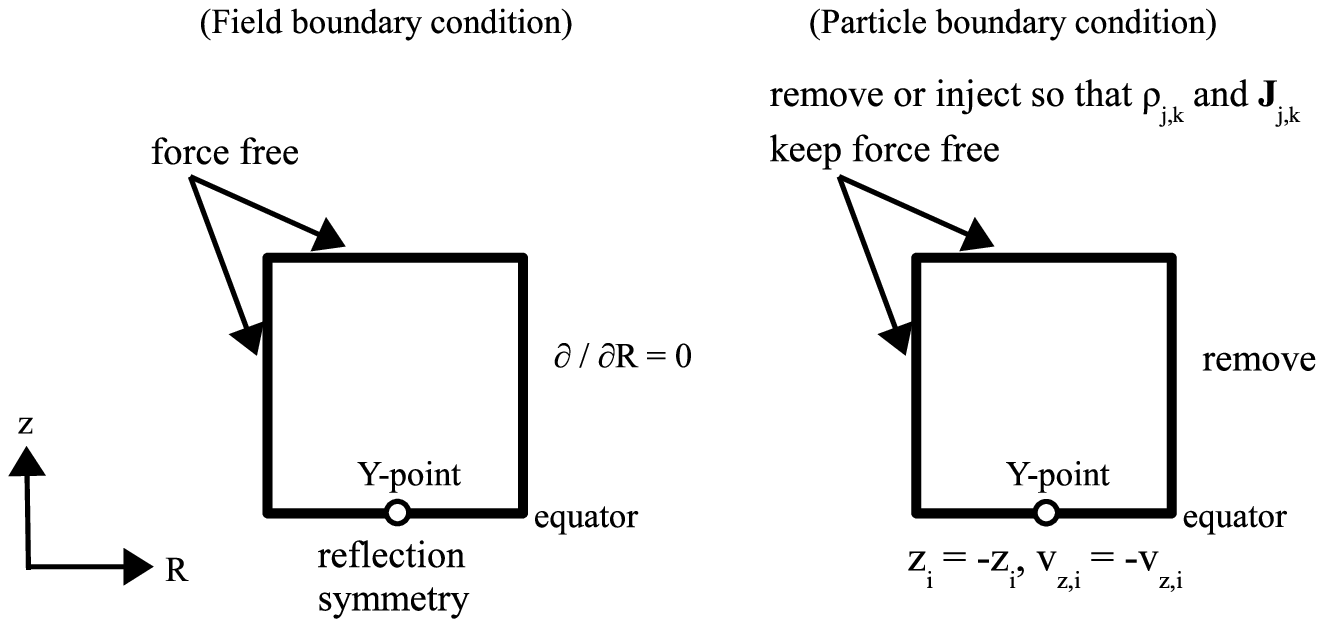}
 \end{center}
 \caption{The boundary conditions for the Y-point simulation.}
 \label{bc}
\end{figure*}

Regarding the local simulation in the global structure shown in figure
\ref{magnetosphere}, the boundary condition is of great importance for the
 simulation of the
Y-point. Inner region would be well approximated by the force-free
solution, so that the left and upper boundaries (inner boundaries) are
assumed to be the force-free solution by \citet{key-2} as shown in
figure \ref{bc}. On the other
hand, it is known that Uzdensky's solution is broken down beyond the
light cylinder. It is likely that the particle inertia becomes
important, and the frozen-in condition breaks down near the Y-point
where the magnetic energy density decreases and plasma is more
important. The outer (right-side) boundary may be crossed by super-fast
wind. Then we impose free boundary for the right-side boundary. The
lower boundary can be treated as reflection symmetry. The boundary
conditions might be fair only if the simulation box is large enough. As
for the initial condition we also use the Uzdensky's solution.
The simulation shall be continued until the system becomes quasi steady state.

\subsubsection{The force-free solution around the Y-point}

As shown by \citet{key-2},
\begin{figure}[h!]
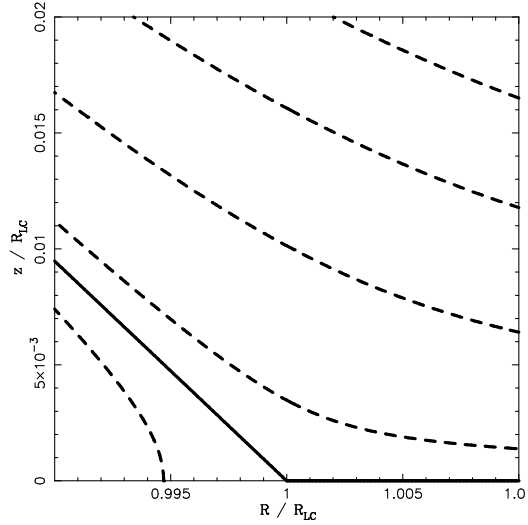

 \begin{center}
  \FigureFile(70.0mm, 70.0mm){./fig14.eps}
 \end{center}
 \caption{The force-free solution around the Y-point given by
 \citet{key-2}. The dashed lines represent poloidal magnetic field
 line. The solid line represents the separatrix, $\Psi=0$, between the open and
 closed field-line regions.}
 \label{solution}
\end{figure}
if the Y-point is located at the light cylinder, the regularity
condition at the light cylinder determines the poloidal magnetic flux
, $\Psi$, and the poloidal current, $\tilde{ I} \left( \Psi \right)$.
The obtained magnetic field is shown in figure \ref{solution}.
Once the force-free solution $\Psi$ and $\tilde{ I} \left( \Psi
\right)$ are obtained, the electromagnetic field, charge density, and
current density are given by
\begin{eqnarray}
 \boldsymbol {B}_{\mathrm{p}} &=& - \frac{\boldsymbol { e}_{\varphi}}{R} \times \nabla
  \Psi, \label{Bp}\\
 {B}_{\varphi} &=& \frac{\tilde { I}}{R},\label{Bphi}\\
 \boldsymbol {E} &=& - \frac{\Omega \left( \Psi \right)}{c} \nabla \Psi\label{E},\\
 \rho &=& \frac{1}{4\pi} \nabla \cdot \boldsymbol { E}, \label{divE}\\
 \boldsymbol { J}_{\mathrm{p}} &=& \frac{c}{4\pi} \frac{d \tilde{I}}{d \Psi} \boldsymbol { B}_{\mathrm{p}}, \label{Jp}\\
 J_{\varphi} &=& -\frac{c}{4\pi R} \left( \nabla^2 \Psi - 2 B_z \right), \label{Jphi}
\end{eqnarray}
where $\Omega$ is the angular velocity of the star.

The bulk velocities of the plasma are restricted by
\begin{equation}
 \boldsymbol { v}_{\pm} = \boldsymbol { \Omega} \times \boldsymbol { r}
  + \kappa_{\pm} \boldsymbol { B}, \label{v}
\end{equation}
where + and - respectively indicate positron and electron, $\boldsymbol
{ r}$ is the position vector, $\boldsymbol{ \Omega}$ is the angular
velocity vector, and $\kappa$ is a scalar function to be determined.

Since the force-free solution does not concern about the plasma density
and bulk velocity, we introduce two model parameters,
\begin{equation}
 n = n_+ + n_-
\end{equation}
and
\begin{equation}
 \bar{\gamma} = \frac{1}{\sqrt{1-(\bar{v}/c)^2}},
\end{equation}
where $\bar{v}= \left| (\gamma_+ n_+ \boldsymbol { v}_+ + \gamma_- n_- \boldsymbol { v}_-) /
 (\gamma_+ n_+ + \gamma_- n_-) \right|$.
Giving the two model parameters, we obtain $n_{\pm}$ and
 $\boldsymbol {v}_{\pm}$, which satisfy the force-free solution for $\rho$
 and $\boldsymbol{ J}$.
 The $N_{j,k}=n_{\pm} V_{j,k}$ particles, whose velocity is
 $\boldsymbol { v}_{\pm}$, are distributed uniformly in the
 ($j$,$k$)-cell by a random number generator.

\subsubsection{Inner boundary condition}

\begin{figure}[h!]
 \begin{center}
  \FigureFile(70.0mm, 70.4mm){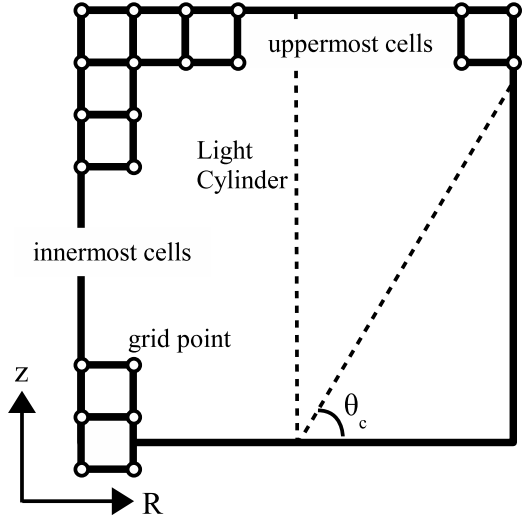}
 \end{center}
 \caption{The innermost cells and the uppermost cells. We imposed
 force-free condition for those cells. In the wedge shape region $\theta
 < \theta_{\mathrm {c}}$, Uzdensky solution breaks down.}
 \label{cell}
\end{figure}

The inner boundary conditions are imposed for the innermost and
uppermost cells (figure \ref{cell}). For the inner boundary cells,
 we assume the density, $n$, so that the number of particle in a cell,
$N_{j,k}=n V_{j,k}$, is constant.
With the force-free charge density, we have the particle numbers for
each species to be distributed in each cell,
\begin{equation}
 N_{\pm}=\frac{1}{2} \left( N_{j,k} \pm \frac{\rho}{q} V_{j,k} \right),
\end{equation}
where we typically take $N_{j,k}$=constant=100.
The field-aligned velocity of the particles are restricted by the
force-free current density, i.e.,
\begin{equation}
 q (n_+ \kappa_+ - n_- \kappa_-) = \frac{c}{4\pi} \frac{d \tilde{I}}{d
  \Psi}, \label{kappa1}
\end{equation}
where the poloidal velocities for each species are given by
\begin{equation}
 \boldsymbol { v}_{\mathrm{p} \pm} = \kappa_{\pm} \boldsymbol { B}_{\mathrm{p}}.
\end{equation}
The mean value of $\kappa$ is given by
\begin{equation}
 \bar{\kappa} = \frac{\gamma_+ n_+ \kappa_+ + \gamma_- n_- \kappa_-}
  {\gamma_+ n_+ + \gamma_- n_-} \label{kappa2}
\end{equation}
where $\gamma_{\pm}=\left[ 1-\left(v_{\pm} / c \right)^2 \right]^{-1/2}$.
Given $\bar{\gamma}$, (\ref{kappa1}) and (\ref{kappa2}) are the
simultaneous equations with respect to $\kappa_{\pm}$. In the present
simulation, we set typically $\bar{\gamma}=50$.
Thus the velocity in the boundary cells are given by (\ref{v}).
The current density $\boldsymbol { J}=q\left( n_+ \boldsymbol { v}_+ -
n_- \boldsymbol { v}_- \right)$ automatically satisfies both (\ref{Jp}) and (\ref{Jphi}).

We remove or inject particles in the innermost and uppermost cells so
 that $\rho_{\pm}$ and $\boldsymbol { J}_{\pm}$ keep these force-free formulae.

\subsubsection{Initialization}

As was mentioned before, the initial state of the plasma is set to be
the Uzdensky's force-free solution. The electromagnetic field is set by
(\ref{Bp}), (\ref{Bphi}) and (\ref{E}). For the particles, we defined
the closed field region, i.e. the dead zone, as $\Psi > 0$ and the open field region, i.e. the wind zone, as $\Psi < 0$.
In the dead zone, we take $N_{j,k}$=100 and $\kappa_{\pm}$=0. It is
known that there is a critical line shown in figure \ref{cell} with the
critical angle $\theta_{\mathrm {c}} \approx 62^{\circ}$ from the equatorial
plane and that in the wedge like region $\theta < \theta_{\mathrm {c}}$,
we have $|\boldsymbol { E}| > |\boldsymbol { B}|$. In the force-free wind
zone, we use the continuity condition, $ \nabla \cdot \left( n_{\pm}
\boldsymbol {v}_{\pm} \right) = 0$, instead of setting $N_{j,k}$, and
take $\bar{\gamma}$ = 50. In the non-force-free wind zone, we assume
$\boldsymbol{ v}_{\mathrm{p} \pm} \parallel \boldsymbol{ B}_{\mathrm{p}}$ instead of
 using (\ref{v})
, and take $N_{j,k}$=100 and $\bar{\gamma}$ = 50.
In any case, the memory of the initial condition will be lost in a few
light-transit times, so that the final state will not be affected by the
initial condition.

\subsubsection{Normalization}

We normalize the distance, velocity, and magnetic field respectively by
 the light radius,
$R_{\mathrm{LC}} = c / \Omega$, the speed of light, and $(B_{\mathrm{*}}/2)( R_{\mathrm{*}} / R_{\mathrm{LC}} )^3$, where $B_{\mathrm{*}}$
is the field strength at the poles and $R_{\mathrm{*}}$ is the radius of the star,
respectively. In order to increase computational efficiency, we
renormalize distance and time by the grid spacing, $\Delta x$, and
 $\Delta t / 2$, respectively. 

Normalization constants and renormalization constants are shown in table
 \ref{tab:norm}.

 \begin{table*}[h!]
  \caption{The normalization constants and renormalization constants.}
  \label{tab:norm}
  \begin{center}
   \begin{tabular}{|l|l|l|}
    \hline
    & Normalization constants 
    & Renormalization constants \\
    \hline
    distance
    & $x_0 = R_{\mathrm{LC}}$
    & $x_{00} = \Delta x$ \\
    \hline
    time
    & $t_0 = R_{\mathrm{LC}} / c$
    & $t_{00} = \Delta t / 2$ \\
    \hline
    velocity
    & $v_0 = c$
    & $v_{00} = \Delta x / (\Delta t / 2)$ \\
    \hline
    electromagnetic field
    & $B_0 = (B_{\mathrm{*}}/2)( R_{\mathrm{*}} / R_{\mathrm{LC}} )^3$ 
    & $B_{00} = \Delta x^{-1/2} / (\Delta t / 2)$ \\
    \hline
    charge density
    & $\rho_0 = B_0 R_{\mathrm{LC}}^{-1}$
    & $\rho_{00} = \Delta x^{-3/2} / (\Delta t / 2)$ \\
    \hline
    current density
    & $J_0 = B_0 R_{\mathrm{LC}}^{-1} c$
    & $J_{00} = \Delta x^{-1/2} (\Delta t / 2)^{-2}$ \\
    \hline
    charge
    & $q_0 = B_0 R_{\mathrm{LC}}^2$
    & $q_{00} = \Delta x^{3/2} / (\Delta t / 2)$ \\
    \hline
    mass
    & $m_0 = B_0^2 R_{\mathrm{LC}}^3 c^{-2}$ 
    & $m_{00} = 1$ \\
    \hline
   \end{tabular}
  \end{center}
 \end{table*}

\subsection{Initial result of the Y-point simulation}

\begin{figure}[h!]
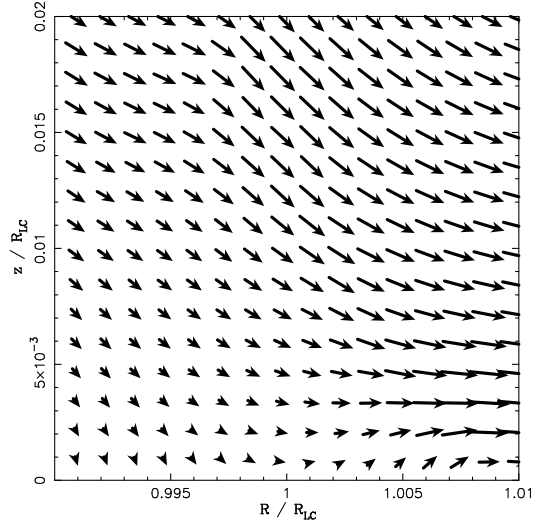

 \begin{center}
  \FigureFile(70.0mm, 70.0mm){./fig16.eps}
 \end{center}
 \caption{The vector of the averaged poloidal magnetic field around the Y-point.}
 \label{yp}
\end{figure}

We have performed a test run for the Y-point. After a
time of $1/60$ rotation, the plasma attains a steady state, in some
sense, with quasi-periodic variability with time scale of $1/300$
rotation.
 Figure \ref{yp} shows the
poloidal magnetic field vector averaged over $1/300$ rotation to indicate the
formation of Y-point around the
intersection of the light cylinder and the equatorial plane. In a short
time scales, however, we see that some part of the closed field line at
the top of the dead zone is broken to make a magnetic island with
magnetic reconnection, and the island go outward. However, sometimes
such a island can come back and merge with the dead zone. Basically,
plasma flows out across the outer (right) boundary. However, the plasma
near the equator move out and in quasi-periodically. In the run,
magnetic reconnection is essential, and the associated heating and
acceleration should be important.  Figure \ref{eb} shows the ratio,
 $|\boldsymbol{E}|$/$|\boldsymbol {B}|$, averaged over the
 same period. The region where $|\boldsymbol{E}| > |\boldsymbol {B}|$
 degenerates near the equatorial plane. It is suggested that the
 electric-field-dominant region becomes several Debye length in
 thickness. With the initial run, we suggest that the Y-point can be
 active in magnetic field dissipation associated by magnetic
 reconnection. The electric-field-dominant region ($|\boldsymbol{E}| >
 |\boldsymbol {B}|$), which is suggested by the force-free solution,
 shrinks into a thin equatorial sheet.

In a subsequent paper, we study the
acceleration and heating, namely the conversion of magnetic energy into
plasma quantitatively.

\begin{figure}[h!]
 \begin{center}
  \FigureFile(70.0mm, 70.0mm){./fig17.eps}
 \end{center}
 \caption{The $|\boldsymbol{E}|/|\boldsymbol{B}|$ distribution around the
 Y-point.}
 \label{eb}
\end{figure}

\section{Summary}

We develop the axisymmetric PIC code in order to analyze
 the Y-point in the pulsar magnetosphere. The code is verified by the
 dispersion relations of waves in electron-positron plasmas.
We also established the appropriate boundary conditions for the Y-point
 in use of the force-free semi-analytic solution. We have performed a
 test run and find that the Y-point is actually formed at the top of the
 dead zone on the light cylinder. It is, however, variable with magnetic
 reconnection. We suggest that the Y-point can be a region around which
 the magnetic field energy convert into the plasma heat and outflow and
 subsequently into radiation. We also suggest that some plasma jet can
 be flow back to polar cap region to produce radio emission. In the next
 step, we will perform large scale simulations and analyze the Y-point
 structure in detail.

\bigskip
Numerical computations were in part carried out on NEC SX-9 at Center
for Computational Astrophysics, CfCA, of National Astronomical
Observatory of Japan and in part carried out on PRIMEPOWER 850
 at Networking and
Computing Service Center of Yamagata University. This work was supported in part by a Grant-in-Aid
from the Ministry of Education, Science, Sports and Culture of Japan
(19540235).
The page charge of this paper is supported by CfCA.

\end{document}